\documentclass[10pt,twocolumn,final]{revtex4}
\usepackage{latexsym}
\usepackage{amssymb}
\usepackage{amsopn}

\newcommand\eq[1]{Eq.~(\ref{#1})}

\newcommand\eqreff[1]{(\ref{#1})}

\newcommand\pa{\partial}

\newcommand\ee{\end{equation}}
\newcommand\be{\begin{equation}}
\newcommand\eea{\end{eqnarray}}
\newcommand\bea{\begin{eqnarray}}

\newcommand\bfk{{\bf k}}

\newcommand\bfx{{\bf x}}

\newcommand\MeV{\,\mbox{MeV}}

\newcommand\sub[1]{_{\rm #1}}
\newcommand\su[1]{^{\rm #1}}

\newcommand\mone{^{-1}}

\newcommand{\vev}[1]{{\langle #1 \rangle }}

\begin{document}
%%%%%%%%%%%%%%%%%%

\title{Classicality of the primordial perturbations}
\author{David H.~Lyth$^1$ and David Seery$^2$}
\address{$^1$ Physics Department, University of Lancaster, Lancaster LA1 4YB, 
United Kingdom}
\vspace{2mm}
\address{$^2$ Astronomy Unit, School of Mathematical Sciences,
Queen Mary, University of London, Mile End Road, London, E1 4NS,
United Kingdom} 
%\date{\today}
\pacs{04.70.-s, 98.80.-k}
%\submitto{JCAP}

\begin{abstract}
We show that during inflation,
a quantum fluctuation becomes classical at all orders if it becomes classical
at first order. Implications are discussed.
\end{abstract}
\maketitle

\section{Introduction}
The promotion of cosmology from an area of speculation
to an area of observation and measurement
is one of the most remarkable developments in the history of
human knowledge.
Starting at an `initial' 
temperature of a few MeV, the observable Universe
is understood in considerable detail (we are setting
$\hbar=c=k\sub B=1$).
At the initial epoch the expanding Universe is
an almost isotropic and homogeneous gas. The perturbations away from
perfect isotropy and homogeneity present at the initial epoch 
are the subject of intense study at present, because they determine
the subsequent evolution of all future perturbations.

According to observation, the dominant initial perturbation is 
the curvature perturbation $\zeta$, so-called because it is related
to the perturbation in the intrinsic curvature of space-time slices with
uniform energy density.
Other initial perturbations may be detected in the future.
The usually-considered examples are a 
perturbation in the composition of the cosmic fluid 
(an isocurvature perturbation), a tensor perturbation setting the initial
amplitude of primordial gravitational waves, and a primordial magnetic field.

To understand the nature and origin of the initial perturbations, 
one should 
use comoving coordinates $\bfx$ that move with expansion of the unperturbed
Universe. After a mapping from the unperturbed Universe to the perturbed one
has been chosen, the perturbations are functions of these 
coordinates. Such a mapping is called a gauge, and indeed is analogous to
the field-theoretic gauge, which also needs to be fixed if gauge fields
play a role.

It is convenient to consider the Fourier components 
with comoving wave-vector $\bfk$. Physical positions are $a(t)\bfx$ and
physical wave-vectors are $\bfk/a(t)$, where $a$ is the scale factor of the 
Universe, and one also defines the Hubble parameter $H=\dot a/a$.

The crucial point now is that gravity slows down the expansion of the
gaseous cosmic fluid,
which means that $aH/k\equiv \dot a/k$ increases as we go back in time.
At the present epoch scales of cosmological interest correspond to
$10^{-6}\lesssim aH/k\lesssim 1$, but at the `initial' temperature $T\sim \MeV$
they all correspond
to $aH/k\gg 1$. Such scales are out of effective causal contact, because
the rate of change of the inverse wavenumber, $\dot a/k$, is bigger than 1.
They are said to be {\em outside the horizon}.

To explain the origin of the perturbations,
it is supposed that going further back in time we reach an era
of inflation
	\footnote
	{The term inflation usually
	denotes an era of expansion corresponding to $\ddot a>0$,
	sometimes described as `repulsive
	gravity'. For the present purpose it might instead be one
	of contraction, corresponding to $\ddot a < 0$,  
	though predictions in that scenario depend on the unknown
	physics of the bounce.},
when by definition
$aH/k$ decreases again to cross the horizon
at the epoch $aH=k$. 
With mild assumptions, it can be shown that inflation drives all perturbations
to zero at the classical level \cite{wald}.
But each bosonic field has a vacuum fluctuation, and one or more of these
fluctuations is supposed to become classical around the time of horizon exit,
the idea being that the timescale of the would-be fluctuation becomes
longer than the Hubble time $H\mone$.
On this basis, the correlators of the classical perturbation are identified 
with quantum expectation values which can be calculated
once one has adopted a specific theory for the inflationary era.
Finally, the classical evolution after horizon exit is supposed to produce
the perturbations at $T\sim \MeV$.

The nature of the evolution after horizon entry is not the 
concern of the present paper. Rather, we want to show that 
the quantum-to-classical transition can actually occur.
This issue has so-far been addressed only in the context of 
first-order calculations, even though higher-order
calculations 
\cite{Maldacena:2002vr,Creminelli:2003iq,Seery:2005wm,Seery:2005gb,%
Weinberg:2005vy,Seery:2006js,Huang:2006eh,sloth} 
have now been done.
We begin by reviewing the first-order case, keeping
the discussion very general so as to include the quite wide variety of 
scenarios that have been proposed.

\section{First order}

Let $\phi_\bfk$ be a Fourier component of any perturbation
existing around the time of horizon exit. (It might be a scalar field,
or else a component of a higher-spin bosonic field or of a 
metric perturbation.)
To first order it will satisfy a linear evolution equation, which 
we take to be of the form
\be
  \phi_\bfk'' + \omega^2_k(k,\eta) \phi_\bfk = 0 \label{phieq}
  ,
\ee 
where a prime denotes $d/d\eta=a(t) d/dt$.
Well before horizon exit the expansion rate is negligible compared with the
physical wavenumber $k/a$, so that are dealing with flat spacetime
and a slowly-varying angular frequency of the form
$\omega_k = c\sub s(k,\eta) k$. Usually $c\sub 
s= 1$ corresponding to a canonically normalized field with negligible
mass, but scenarios \cite{Armendariz-Picon:1999rj,Seery:2005wm} 
exist with $c_s \ll 1$.
In the latter case the epoch of horizon exit 
(when the quantum-to-classical transition occurs) should be redefined as 
$c_s k\sim aH$.

In writing \eq{phieq} we suppose that the perturbations can be chosen 
so that there are no linear couplings between them, which would make the
right hand side a linear combination of the other perturbations.
This will usually be a good approximation as we deal only with the few
Hubble times around horizon exit. The inclusion of linear couplings makes
no difference, 
provided that they are negligible well before horizon exit, as they certainly
are in the usual case corresponding to $c_s=1$.

In the Heisenberg picture the corresponding operator $\hat\phi_\bfk$
has the classical  time-dependence, and the reality condition
  $  \hat\phi_{-\bfk} =\hat\phi_\bfk^\dagger
  $
determines its form as
  $  \hat\phi_\bfk(\eta) = v_k^*(\eta) \hat a_\bfk + 
     v_k(\eta) \hat a^\dagger_{-\bfk}
  $,
where the mode function $v_k(\eta)$ also satisfies \eq{phieq}. 
To determine the quantum theory we need only consider the epoch
well before horizon exit. Starting from the action for $\phi_\bfk$
we apply canonical quantization, which after fixing the normalization of
the commutator
  $  [\hat a_\bfk,\hat a^\dagger_{\bfk'}]\propto \delta(\bfk-\bfk')
  $
determines the Wronskian  $v'_k v_k^*-v_k v_k^*{}'$ of the mode function. 
Then we choose a particular mode function, which is usually taken to be
\cite{Birrell:1982ix,Mukhanov:2005sc}
be
  $  v_k\propto e^{ \pm ic\sub s k\eta}
  $.
(The $\pm$ choices are physically equivalent; a mixture
  $  v_k=\alpha_k e^{+ ic\sub s k\eta} + \beta_k  e^{- ic\sub s k\eta}
  $
is sometimes considered \cite{bm}.)
Next we use the creation operator
$\hat a_\bfk^\dagger$ to construct the Hilbert space (Fock space),
starting from the vacuum annihilated by $\hat a_\bfk$. A similar Hilbert
space is constructed for every component of every field. The complete
Hilbert space is the direct product of these,
together with another Hilbert space describing
the fields which do not correspond to perturbations and are irrelevant in
the free-field approximation. Finally, the state vector is specified, which
describes our Universe well before horizon entry.  
This setup defines, at any instant in time, an ensemble of universes specified
by different values of $\phi_\bfk$. The ensemble corresponds to a gaussian
random field, meaning that the two-point correlator 
$\vev{\hat\phi_\bfk\hat\phi_{\bfk'}}$ is the only connected one. 

Now comes the point. We make the usual assumption that the 
state vector corresponds to the vacuum, for at least \cite{abook,dl}
most of the states
	\footnote
	{An alternative is Warm Inflation,
	where one supposes that the 
	cosmic fluid has a small thermalized component whose classical 
	thermal fluctuation generates the curvature perturbation by freezing 
	out at horizon exit \cite{warm}.
	Then the curvature 
	perturbation is classical even before horizon
	exit, but the present discussion is still relevant for 
	possible perturbations other than the curvature perturbation.}.
Then the fluctuation
$\phi_\bfk$ {\em well before horizon entry}
is definitely a quantum object; the wave
function giving its probability distribution is so narrowly 
peaked that no wave-packet state vectors exist, such that
$\phi_\bfk(t)$ has a well defined value.
(We are talking about values with reasonable probability of course, as opposed
to large values far out on the tail of the probability distribution.
Our universe is supposed to correspond to a typical member of the ensemble,
and therefore the perturbations we measure are not supposed to deviate
too much from the ensemble average.)

One has to show that wave-packet state vectors
{\em do} exist after horizon
exit. This will be the case if
the phase of the mode function can be chosen so that it becomes real in the 
limit $k/aH\to 0$, so that \cite{starob82,starob86}
\be
  \hat\phi_\bfk(\eta)  \to v_k(\eta)  \hat A_\bfk
  \label{real} ,
\ee
where $\hat A_\bfk$ is a constant operator (actually equal to $\hat a_\bfk
+\hat a_{-\bfk}^\dagger$).
Indeed, in this limit a state vector which is an eigenvector
of $\hat\phi_\bfk$ at any instant 
remains an eigenvector
	\footnote
	{Remember that all state vectors are time-independent in the Heisenberg
	picture. A derivation of classicality in the Schr\"{o}dinger picture
	was given in \cite{lyth85,gp} and its equivalence to the Heisenberg picture
	is discussed in \cite{Polarski:1995jg}.}.

We need the condition \eqreff{real} to be well-satisfied
before the end of inflation, which we assume is sufficient for the
perturbation to behave classically long after horizon exit.
One might object that the fact the limit is never actually achieved
makes this formalism non-rigorous, but the deep problems of
interpretation which afflict quantum theory in a cosmological context
mean that the critera for exact classical behaviour are
not accessible in our present state of knowledge.
In any case,
it seems unavoidable that \eqreff{real} will be a necessary
pre-requisite for classicality in any more complete
theory that describes quantum processes in the Universe.

We will refer to fields satisfying \eq{real} as light fields.
This terminology is motivated by
the example of a free scalar field with the canonical kinetic term and
mass $m$, living in unperturbed de Sitter space
corresponding to constant $H$. In that case the mode function becomes real
if $m^2<(9/4)H^2$.
Assuming general relativity, the tensor metric perturbation 
has the dynamics of a massless 
canonically-normalized scalar field, and has little effect on the 
expansion so that it will certainly be a light field.
A massless
spin-1 field with minimal kinetic term is {\em not} light in our sense
because its contribution to the action is invariant
under a conformal transformation to flat spacetime, which means that its
mode function continues to oscillate after horizon exit. Such a field
will be light if it has a 
sufficiently small and slowly-varying mass (generated by a Higgs mechanism
and/or a suitable non-minimal kinetic term), which nevertheless is big enough
for \eq{real} to be satisfied before the end of inflation.
A gauge field of this sort
might create a primordial magnetic field, or have
more dramatic \cite{kdimop} effects. 
These are the only types of bosonic field that are usually
considered in the context of quantum field theory, 
but if others exist our treatment will cover them. 

The condition \eqreff{real} is sometimes called WKB classicality.
Once it is satisfied one might worry about the
`Schr\"{o}dinger's Cat' problem of how the initial state collapses into a
particular state with definite $\phi_\bfk$. There is also the issue of
decoherence, which addresses the question of why definite 
values of $\phi_\bfk$ are preferred (why these are `pointer states').

Our aim though, is to show that WKB classicality occurs also in the 
presence of interactions. In preparation for that, we write \eq{real}
in the equivalent form
\be
  [\hat\phi_\bfk,\hat\phi_\bfk'] \to 0
  \label{class}
  .
\ee
Indeed, \eq{real} obviously implies \eq{class}. Conversely,
\eq{class} implies that
eigenvectors of $\hat \phi_\bfk$ become also eigenvectors of $\hat\phi_\bfk'$
because the eigenvalues of $\hat\phi_\bfk$ are non-degenerate.
Thus \eq{class} implies \eq{real}, establishing their equivalence.
We note that the field $\hat\phi_\bfk$ which appears in \eq{class}
is chosen to be in the Heisenberg picture, in which state vectors are
time-independent while the fields evolve. Therefore,
this is the appropriate classicality condition for both
free and interacting fields.

\section{Higher order}

At first order the only connected correlator is the two-point one, so that 
the perturbations are gaussian. Non-gaussianity 
will be generated at higher order, and may 
be a valuable discriminator between models. For this reason second order
\cite{Maldacena:2002vr,Creminelli:2003iq,Seery:2005wm,Seery:2005gb}
and third order \cite{Seery:2006js,Huang:2006eh} calculations have been done,
yielding estimates of respectively three-point 
and connected four-point correlators.

Higher order effects will come from
ordinary interactions that exist even in flat spacetime, and from
the gravitation theory which determines the metric perturbations.
To handle them we can write $\hat H=\hat H_0+\hat H_I$, 
and work in the interaction picture so
that the evolution of the  perturbations is  given by
  $  \hat\phi_\bfk'=i[\hat H_0,\hat\phi_\bfk] \label{dphidt}
  $.
The term $\hat H_I$ involves all of the fields, bosonic and
fermionic, in the curved spacetime quantum field theory that is being invoked.
From now on we drop the subscript $\bfk$.

An operator $U$ transforms each 
operator $A$ to the Heisenberg picture through 
  $  A\sub{hp}(t) = U\mone(t) A(t) U(t)
  $.
The form of $U$ is determined by
  $  A\sub{hp}' = i [ H\sub{hp}, A\sub{hp} ]
  $,
giving
  $  \dot U(t) =  -i \hat H_I(t) U(t)
  $,
whose solution is
\be
  U(t) = T \exp \left( -i \int^t_{-\infty} \hat H_I(t') dt' \right)
  \label{U}
\ee
where $T$ is the time-ordering operator. 

Eq.~(\ref{U}) has the formal power series expansion
\be
  U(t) = 1+ \sum_{n=1}^\infty \frac{(-i)^n}{n!}
  \int dt' \cdots dt'' \; T \left[
    \hat H_I(t') \cdots \hat H_I(t'') \right] , \nonumber
\ee
which can be truncated at any required order $n$ 
when $\hat H_I$ is perturbatively small. 
The vacuum expectation values of products of the  
perturbations at a fixed time (correlators) can then be calculated
to this order, using
the Schwinger formalism, as described in
\cite{Maldacena:2002vr,Weinberg:2005vy,Calzetta:1986ey,Jordan:1986ug}
which is known as the `in--in' or `closed time path' formalism.
Each correlator is given by a sum of Feynman diagrams 
\cite{Weinberg:2005vy}.

From the discussion after \eq{class}, we see that WKB classicality requires
$[\hat\phi\sub{hp},\hat\phi\sub{hp}']\to 0$.
Written in terms of interaction picture quantities this requirement becomes
$U B  U\mone \to 0$, where
\be
  B \equiv   i[\hat\phi,[\hat H_I,\hat\phi]] + [\hat\phi,\hat\phi'] 
  .
\ee
The second term of $B$ tends to zero in accordance with \eq{class}.
The first term is zero if $H_I$ does not involve $\hat\phi'$,
since $\hat\phi$ commutes at equal times with every other interaction-picture
field. If $H_I$ does contain derivatives of the fields (which
is the case in almost all interesting examples), then the first term
will also approach zero as \eq{class} becomes increasingly well-satisfied.

Since $B$ is approaching zero,
the WKB classicality requirement will therefore be satisfied if $U\simeq 1$
to sufficient accuracy. This will be the case whenever perturbation theory
applies, and could even be the case in a mildly non-perturbative situation.
The requirement $U\simeq 1$ is of course essential, since complete freedom
to choose $U$ would allow us to choose $\hat H_I = \hat H\sub{new} - \hat H_0$
for any $\hat H\sub{new}$, leading to the wrong conclusion that all quantum
fluctuations become classical after horizon exit. As we noticed earlier,
a very massive scalar field fluctuation does not become classical, and neither
does a massless vector field fluctuation. It is actually safest to include
all mass terms in $H_0$, though that is not compulsory in the cosmological
context because well before horizon exit they become negligible anyway.

\section{Discussion}

It is instructive to consider the implications of our result in a definite
setting, which we take to be the one
described in
\cite{Maldacena:2002vr,Seery:2005gb,Weinberg:2005vy}.
The action is Einstein--Hilbert with canonical kinetic terms for the light
scalar fields. Around the time of horizon exit for the chosen scale, 
slow-roll inflation occurs in the direction of the inflaton
$\phi$, and any orthogonal light fields $\sigma_i$ have no effect at that time.
The light scalar field perturbations may be defined on `flat'
spacetime slices of fixed $t$, defined as those whose metric is of the form
  $  g_{ij} = a^2(t) [ \exp(\tilde h) ]_{ij}
  $,
with $\tilde h(\bfx,t)$  transverse and traceless.
Instead of the flat slicing one can choose a fixed-$t$
slice of uniform energy density, writing
  $  g_{ij} = a^2(t) \exp(2 \zeta) [\exp(h)]_{ij}
  $, 
with $h(\bfx,t)$ transverse and traceless. Then $\delta\phi$ vanishes
and the curvature perturbation  $\zeta(\bfx,t)$ 
becomes the degree of freedom.

Now consider an $n$-point correlator
  $  \langle \delta\phi_{\bfk_1}
     \cdots \delta\phi_{\bfk_n} \rangle
  $
of light fields.
Once the condition \eq{class} becomes well-satisfied,
our result implies that such a correlation function can be computed
by spatially averaging an
effective classical field $\delta\phi\su{cl}_{\bfk}$.
It is now possible to make an immediate application to
a calculation of Weinberg \cite{Weinberg:2006ac}.
He finds,
under stated assumptions but to all orders in perturbation theory,
that no correlator of light field perturbations
increases as fast as a power of the scale factor.
Weinberg's argument can actually be applied directly to the effective
classical field $\delta\phi\su{cl}_{\bfk}$. 
To be precise, the quantity $Q$ which appears in
Eq.~(7) of \cite{Weinberg:2006ac} 
can refer to the time dependence of $\delta\phi\su{cl}$
rather than merely its expectation value. Because the correlators are
obtained by spatially averaging the effective field, there is no
need to consider $Q$s formed from a product of perturbations.

Now we come to a more general consideration.
Suppose that the values of some set of light scalar fields $\phi_i(\bfx)$,
evaluated at some epoch during inflation
and smoothed on a scale bigger than $H\mone$ so that they are classical,
determine the future evolution of the locally-defined
scale factor $a(t)\exp[\zeta(\bfx,t)]$
at each position $\bfx$. Then \cite{Lyth:2006gd,lcurv}
\be
  \zeta(\bfx,t)  = \sum_i N_i(t) \delta\phi_i(\bfx) 
  + \sum_{ij} \frac{1}{2} N_{ij}(t)  \delta\phi_i(\bfx)
    \delta\phi_j(\bfx) + \cdots
  \label{zeta} 
  ,
\ee
where $N_i\equiv \pa N(\phi_i,\rho(t))/\pa \phi_i$ etc., and $N$ is 
the number of $e$-folds evaluated in a family of 
unperturbed universes, starting from an epoch when the light fields have 
assigned values and ending when the energy density has an assigned value.

This expression gives the correlators of $\zeta$ in terms of the correlators
of the light fields, which in turn are given by the 
in--in Feynman diagrams of quantum field theory.
Let us call those Q-Feynman diagrams,
since they originate from the quantum Schwinger formalism.
But the evaluation of the correlators
from \eq{zeta} can itself be 
represented \cite{bu,Boubekeur:2005fj,lcurv,sloth} by Feynman
diagrams involving integrations over 
products of Fourier components of $\delta\phi_i$. These are in the 
classical regime by virtue of the smoothing, so we refer to them as
C-Feynman diagrams.
The  curvature perturbation, then,  can be obtained by combining Q-Feynman
diagrams referring to the flat slicing, with the C-Feynman diagrams
that take us from the flat slicing to the uniform-density slicing on which
$\zeta$ is defined.

On the other hand, $\zeta$ {\em during inflation} can instead be calculated
directly on the uniform-density 
slicing, which can be represented by its own set of
Q-Feynman diagrams. We conclude that the latter set is equivalent to the
combination of  Q-Feynman and C-Feynman diagrams described in the previous
paragraph. A specific example of this equivalence is provided by the 
tree-level calculation of the three-point correlator of $\zeta$
in the single-field slow-roll inflation, which has
been done on both the uniform-density slicing \cite{Maldacena:2002vr}
 and on the flat slicing \cite{Seery:2005gb} to obtain the same result.

Regarding the equivalence, 
it is important to realise that the epithet `classical'
refers
{\em only} to the time evolution. Despite the epithet, each light field
perturbation $\delta\phi_i$ vanishes in the limit $\hbar\to 0$,
even after horizon exit,
because it originates as a vacuum fluctuation.
Therefore, our result is not inconsistent with well-understood
effects such as quantum particle creation. However, it places non-trivial
constraints on their subsequent evolution once $k/aH \ll 1$.
The same is true for the C-Feynman diagrams, in which each loop
introduces a factor $\hbar$, just like each
loop of an Q-Feynman diagram. This is why the tree-level calculations mentioned
earlier are equivalent; they both are of first order in an expansion in
powers of $\hbar$. It remains to be seen how this type of argument goes
for more general calculations, involving higher correlators and
loop contributions, for which the evolution equation for the effective
classical
field $\delta\phi\su{cl}$ will need to be understood in greater detail. 

%%%%%%%%%%%%%%%%
\section*{Acknowledgments}
%%%%%%%%%%%%%%%%
DS is supported by PPARC grant
PPA/G/S/2003/00076. DHL is  supported by  
PP/D000394/1   and by EU grants  
MRTN-CT-2004-503369 and MRTN-CT-2006-035863. The work was begun in Lancaster
at the workshop \emph{Non-Gaussianity from Inflation} 5--9 June 2006.

\section*{References}
%%%%%%%%%%%%%%%%%%%%%%%%%%%%%%%%%%%%%%%%%%%%%%%%%%%%%%%%%%%%%%%%%%%%

%%%%%%%%%%%%%%%%%%%%%

%%%%%%%%%%%%%%%%%%%%%%%%%%%%%%%%%%%%%%%%%%%%%%%%%
\end{document}